\documentclass[12pt, twocolumn, nofootinbib]{revtex4}
\usepackage{graphicx, multirow, lscape}
\usepackage{amssymb}
\usepackage{setspace}

\begin{document}
\title{\Large{Correlations and clustering in the trading of members of the London Stock Exchange}}
\author{\large{Ilija I. Zovko}} \affiliation{Center for Nonlinear Dynamics in Economics and Finance, University of Amsterdam, The Netherlands\footnote{This work was funded by Barclays Bank and National Science Foundation grant 0624351.  Any opinions, findings, and conclusions or recommendations expressed in this material are those of the authors and do not necessarily reflect the views of the National Science Foundation.}} \email{zovko@santafe.edu}
\affiliation{Santa Fe Institute, 1399 Hyde Park Road, Santa Fe, NM 87501, USA}
\affiliation{RiskData, 6 rue d'Amiral Coligny, 75001 Paris, France}

\author{\large{J. Doyne Farmer}}\affiliation{Santa Fe Institute, 1399 Hyde Park Road, Santa Fe, NM 87501, USA}

\singlespacing

\begin{abstract}
This paper analyzes correlations in patterns of trading of different members of the London Stock Exchange.  The collection of strategies associated with a member institution is defined by the sequence of signs of net volume traded by that institution in hour intervals.  Using several methods we show that there are significant and persistent correlations between institutions.  In addition, the correlations are structured into correlated and anti-correlated groups.  Clustering techniques using the correlations as a distance metric reveal a meaningful clustering structure with two groups of institutions trading in opposite directions.
\end{abstract}

\maketitle

\tableofcontents

\section{Introduction}
The aim of this paper is to examine the heterogeneity of the trading strategies associated with different members of the London Stock Exchange (LSE).  This is made possible by a dataset that includes codes identifying which member of the exchange placed each order.  While we don't know who the member actually is, we can link together the trading orders placed by the same member.  Member firms can be large investment banks, in which case the order-flow associated with the code will be an aggregation of various strategies used by the bank and its clients, or at the other extreme it can be a single hedge fund.  Thus, while we cannot identify patterns of trading at the level of individual trading strategies, we can test to see whether there is heterogeneity in the collections of strategies associated with different members of the exchange.   For convenience we will refer to the collection of strategies associated with a given member of the exchange as simply {\it a strategy} and the member of the exchange as simply an {\it institution}, but it should be borne in mind throughout that ``a strategy" is typically a collection of strategies, which may reflect the actions of several different institutions, and thus may internally be quite heterogeneous.

We define a strategy by its actions, i.e. by the net trading of an institution as a function of time.  If the net volume traded by an institution in a period of time is positive there is an net imbalance of buying, and conversely, if the net volume is negative there is a net imbalance of selling.  We test to see whether two strategies are similar in terms of their correlations in the times when they are net buyers and when they are net sellers.

Two studies similar in spirit to this one are~\cite{Lillo07, Vaglica07} in which the authors analyse trading strategies using data from the Spanish Stock Exchange.  A number of related studies analysing market correlations can be found in~\cite{Laloux99,Laloux99a,Plerou02a,Plerou99a,Potters05, Bonanno00,Mantegna99a,Onnela03}.

\subsection{The LSE dataset}
The LSE is a hybrid market with two trading mechanisms operating in parallel.  One is called the on-book or ``downstairs" market and operates as an anonymous electronic order book employing the standard continuous double auction.  The other is called the off-book or ``upstairs" market and is a bilateral exchange where trades are arranged via telephone.  We analyse the two markets separately.

The market is open from 8:00 to 16:30, but for this analysis we discarded data from the first hour (8:00 -- 9:00) and  last half hour of trading (16:00 -- 16:30) in order to avoid possible opening and closing effects.  

We base the analysis on four stocks, Vodafone Group (VOD, telecomunications), AstraZeneca (AZN, pharmaceuticals), Lloyds TSB (LLOY, insurance) and Anglo American (AAL, mining).  We chose VOD as it is one of the  most liquid stocks on the LSE.  LLOY and AZN are examples of frequently traded liquid stocks, and AAL is a low volume illiquid stock.


%


\subsection{Measuring correlations between strategies}
The institution codes we use in this analysis are re-scrambled by the exchange each month for privacy reasons\footnote{ However, we have found a way to track institutions' trading on the on-book market for longer time periods. We use this fact in a subsequent part of the paper.  More about this later in the text.}.  This naturally divides the dataset into monthly intervals which we treat as independent samples.  The data spans from September 1998 to May 2001, so we have 32 samples for each stock.

In order to define the trading strategies we further divide the monthly samples into hourly intervals.  We believe that one hour is a reasonable choice, capturing short time scale intraday variations, but also providing some averaging to reduce noise.  For each of these hour intervals and for each institution individually, we calculate the net traded volume in monetary units (British Pounds).  Net volume is total buy volume minus total sell volume.  We then assign to each institution and hour interval a +1, -1 or 0 describing it's strategy in that interval.  If the net volume in an interval is positive (the institution in that period is a net buyer) we assign it the value +1.  If the institution's net volume in the interval is negative (the institution is a net seller) we assign it the value -1.  If the institution is not active within the interval we assign it the value 0. We discard institutions that are not active for more than 1/3 of the time. 

\begin{figure*}[htbp]
\begin{center}
\includegraphics[scale=0.6]{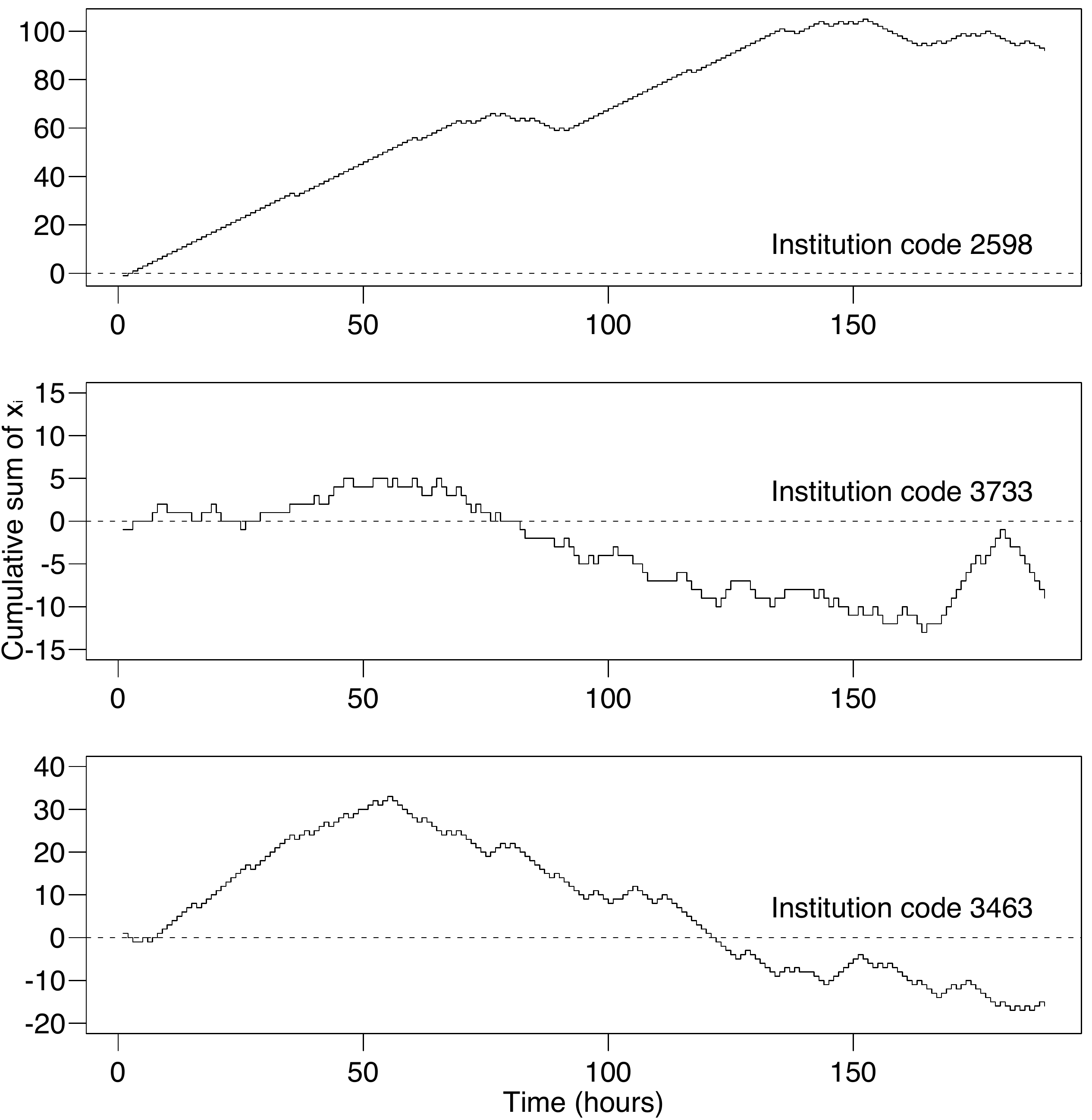}
\caption{Three examples of institutions' strategies for on-book trading in Vodafone.  We plot the cumulative sum of the (+1 , -1 and 0) indicators of hourly net trading volume within a month.  The three institutions were not chosen randomly but rather to illustrate three very different trading styles.  Institution 2598 appears to be building up a position, institution 3733 could be acting as a market maker, while 3463 seems to be a mix of the two.  In reality, only a small number of institutions show strong  autocorrelations in their strategy (such as the top and bottom institutions in the plot) and do not have such suggestive cumulative plots. }
\label{activityMatrix}
\end{center}
\end{figure*}

Three examples of trading strategies are shown in figure~\ref{activityMatrix}.  The examples show cumulated trading strategies for three institutions trading Vodafone on-book in the month of November 2000.

In the end we obtain for each month of trading a set of time series representing the net trading direction for each institution $x_i(t)$, which can be organized in a 'strategy matrix' $M$ with dimensions $N \times T$, where $N$ is the number of active institutions and $T$ is the number of hour intervals in that month.  The number of active institutions varies monthly and between stocks.  Typical value of $N$ for the on-book market for liquid stocks is around $N \sim 70$, and for less liquid stocks  $N\sim 40$.  For the off-book market, the numbers are 1.5 to 2 times larger.  The number of hourly intervals depends on the number of working days in a month, and is around $T=7 \times 20 = 140$.

\begin{figure*}[hptb] 
\begin{center} 
	\includegraphics[scale=0.6]{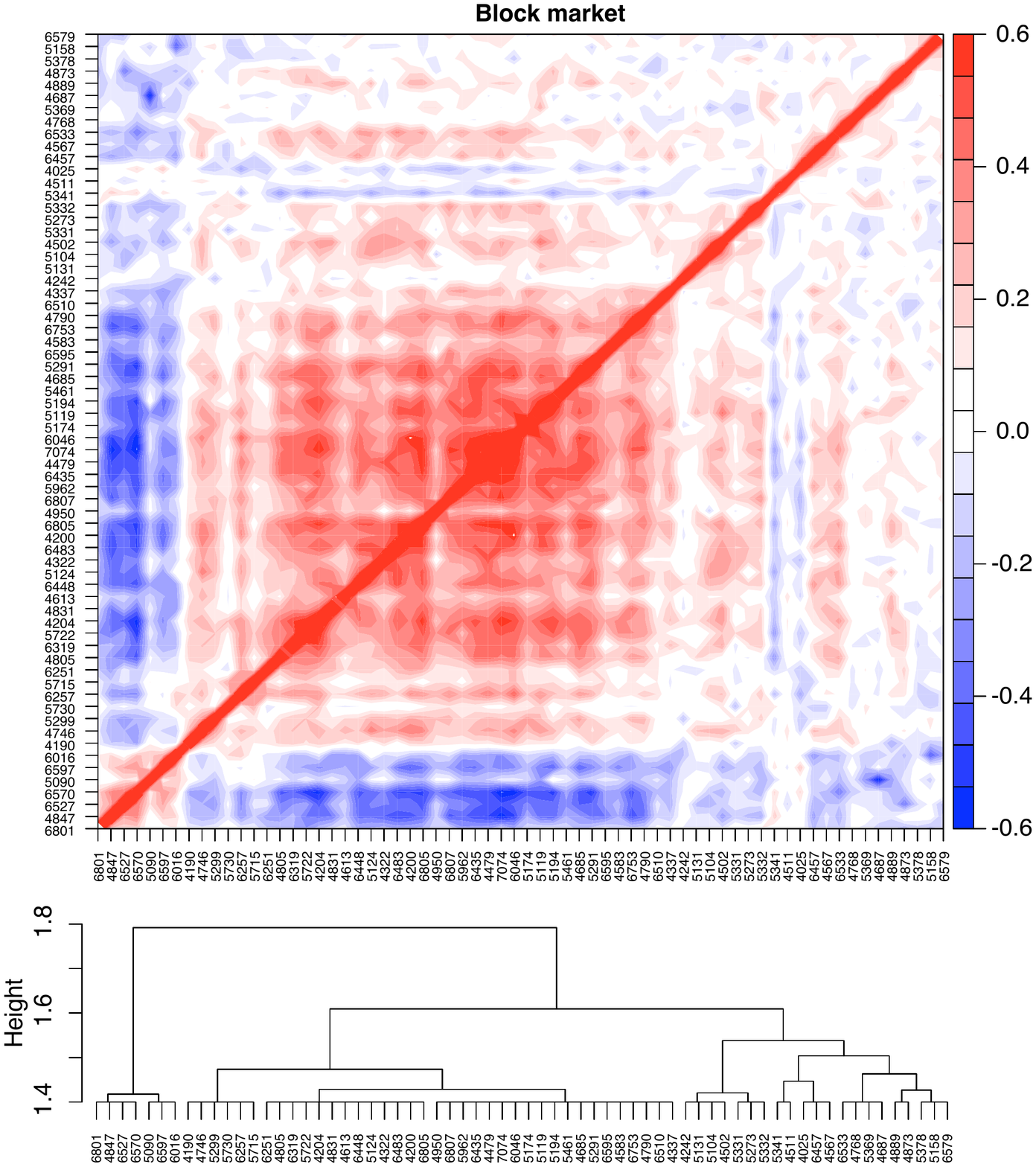}
	\caption{Example of a rearranged correlation matrix for off-book market trading in Vodafone in November 2000.  The ordering of institutions is based on the result of the clustering algorithm explained in section~\ref{clustSect}.  Red colors represent positive correlations between institutions' strategies, blue represent negative correlations, and darker colors are larger correlations.  The dendrogram resulting from the clustering is shown below the correlation matrix.  For visual clarity the cluster is cut at height 1.4.}
\label{dendrograms}
\end{center}
\end{figure*}

Given the monthly strategy matrices $M$ we then construct the $N \times N$ monthly correlation\footnote{Since the data assumes only three distinct values (0,1 and -1) Pearson and Spearman correlations are equivalent.}  matrices between the institutions' strategies.  
A color example of a correlation matrix for off-book trading in Vodafone in November 2000 is given in the top panel of figure~\ref{dendrograms}.  Dark colors represent high absolute correlations, with red positive and blue negative.  Since the ordering of institutions is arbitrary we use the ordering suggested by a clustering algorithm as explained later in the text.  It is visually suggestive that the correlations are not random: Some groups of institutions are strongly anticorrelated with the rest while in turn being positively correlated among themselves. 

A formal test of significance involving the t-test cannot be used as it assumes normally distributed disturbances, whereas we have discrete ternary values.  Later in the text we use a bootstrap approach to test the significance.  Now, however, we test the significance of the correlation coefficients using a standard algorithm as in ref.~\cite{Best75}.  The algorithm calculates the approximate tail probabilities for Spearman's correlation coefficient $\rho$.  Its precision unfortunately degrades when there are ties in the data, which is the case here.  With this caveat in mind, as a preliminary test, we find that, for example, for on-book trading in Vodafone for the month of May 2000, 10.3\% of all correlation coefficients are  significant at the 5\% level. Averaging over all stocks and months, the average percentage of significant coefficients for on-book trading is $10.5\%  \pm 0.4\%$, while for off-book trading it is $20.7\% \pm 1.7\%$.  Both of these averages are substantially higher than the 5\% we would expect randomly with a 5\% acceptance level of the test.

\section{Significance and structure in the correlation matrices}
The preliminary result of the previous section that some correlation coefficients are non-random is further corroborated by testing for non-random structure in the correlation matrices.   The hypothesis that there is structure in the correlation matrices contains the weaker hypothesis that some coefficients are statistically significant.   

The test for structure in the matrices would involve multiple joint tests for the significance of the coefficients.   An alternative method, however, is to examine the eigenvalue spectrum of the correlation matrices.  Intuitively, one can understand the relation between the two tests by remembering that eigenvalues $\lambda$ are roots of the characteristic equation $\det(A-\lambda 1) = 0$, and that the determinant is a sum of permutations of products of the matrix elements $\det(A) = \sum_{\pi} \epsilon_{\pi} \Pi_\pi a_\pi$, where $\pi$ are the permutations and $\epsilon_\pi$ is the Levi-Civita antisymmetric tensor.  On the other hand the test is  directly related to principal component analysis, as the eigenvalues of the correlation matrix determine the principal components.  

The existence of empirical eigenvalues larger than the values expected from the null implies that  there is structure in the correlation matrices and the coefficients are significant.

\subsection{Density of the correlation matrix eigenvalue distribution}
For a set of infinite length uncorrelated time series all eigenvalues of the corresponding correlation matrix (which in this case would be diagonal) are equal to 1.  For finite length time series, however, even if the underlying generating processes are completely uncorrelated, the eigenvalues will not exactly be equal to one - there will be some scatter around one.  This scattering is described by a result from random matrix theory~\cite{Laloux99,Laloux99a,Potters05}.  For $N$ random uncorrelated variables, each of length $T$, in the limit $T \rightarrow \infty$ and $N \rightarrow \infty$ while keeping the ratio $Q=T/N \geq 1$ fixed, 
the density of eigenvalues $p(\lambda)$ of the correlation matrix is given by the functional form 
\begin{eqnarray} 
p(\lambda) &=& \frac{Q}{2 \pi \sigma^2} \frac{\sqrt{(\lambda_{\small max} - \lambda) (\lambda - \lambda_{\small min})}}{\lambda} \nonumber \\ 
\lambda^{\small max}_{\small min} &=& \sigma^2 ( 1+1/Q \pm 2 \sqrt{1/Q}).
\label{eigenDens}
\end{eqnarray} 
$\sigma^2$ is the variance of the time series and $\lambda \in [\lambda_{\small min}, \lambda_{\small max}]$.   Apart from being a limiting result, this expression is derived for Gaussian series.  As it turns out the Gaussian assumption is not critical, at least not for the right limit $\lambda^{\small max}$, which is the one of interest for this study.  We show in a subsequent section a simulation result confirming this observation.


\begin{figure*}[hptb]
\begin{center}
	\includegraphics[scale=0.53]{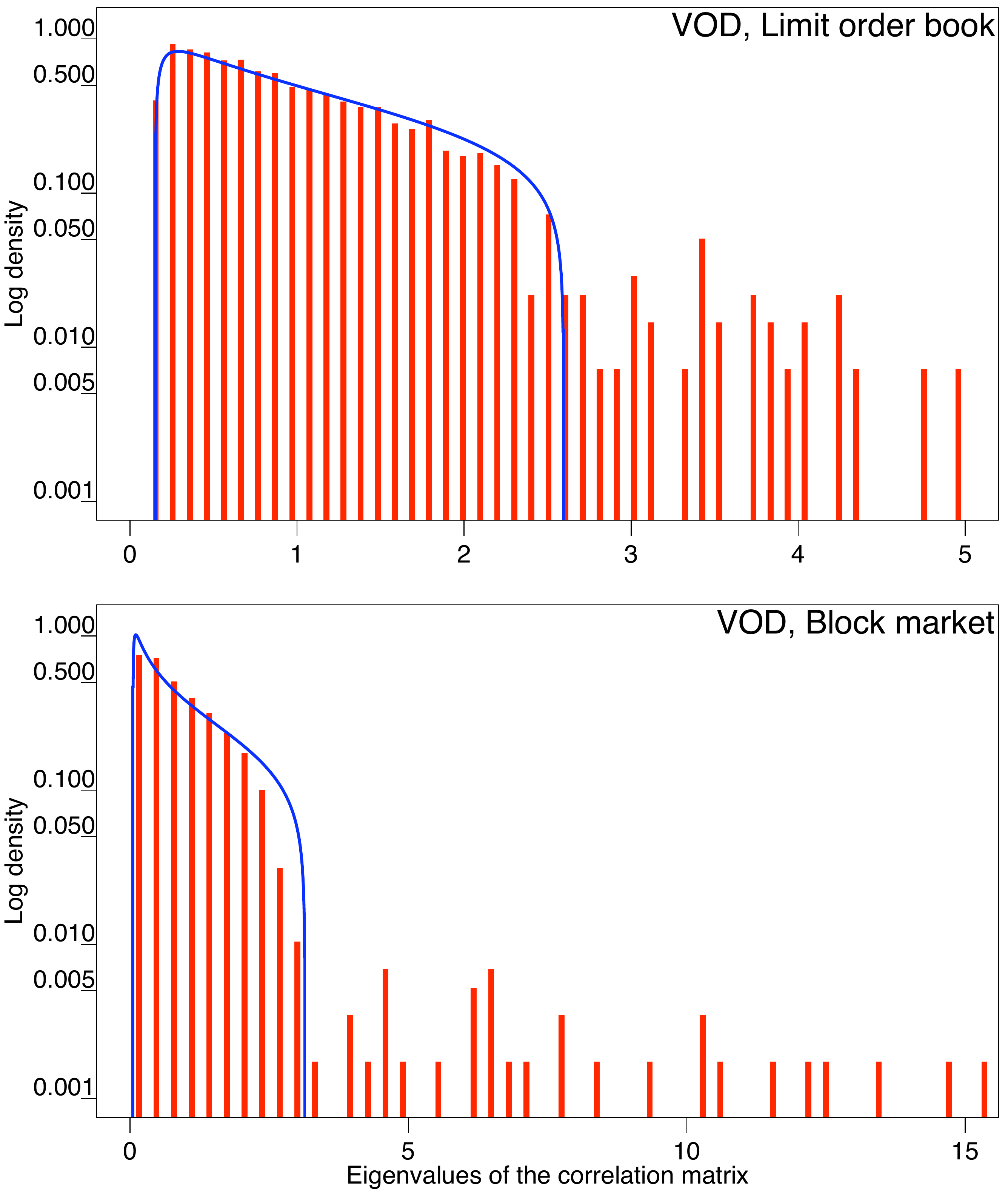}
	\caption{Empirical density of eigenvalues of the correlation matrices (red) compared to the theoretical density for a random matrix (blue).  We see that there are many eigenvalues not consistent with the hypothesis of a random matrix.}
\label{eigenDensityVOD}
\end{center}
\end{figure*}

A further consideration is the fact that the parameters $Q=T/N$ and $\sigma$ change every month\footnote{$\sigma$ is calculated mechanically using the standard formula, as if the time series had a continuous density function rather than discrete ternary values.}, as both the number of hour intervals and the number of institutions vary.  Consequently the predicted eigenvalue density under the null changes from month to month.  In principle we should construct a separate test for each month, comparing the eigenvalues of a particular month with the null distribution using the appropriate value of $Q=T/N$ and $\sigma$.  However, monthly $Q$ and $\sigma$ do not vary too much, and the variation does not change the functional form of the null hypothesis substantially.  In view of the fact that Eq.~\ref{eigenDens} is valid only in the limit in any case, we pool eigenvalues for all months together, construct a density estimate and compare it with the null using the monthly averages of $Q$ and $\sigma$.


Figure~\ref{eigenDensityVOD} shows the empirical eigenvalue density compared with the expected density under the null for the stock Vodafone.  The top figure shows on-book trading, while the bottom figure shows off-book trading.  In both markets there are a number of eigenvalues larger than the cutoff $\lambda^{max}$ and not consistent with uncorrelated time-series.  The eigenvalues are larger in the off-book market because the correlation matrices are larger (there are more traders active in the off-book market than the on-book market).  The largest eigenvalue in the off-book market is 5 times the noise cutoff whereas it is only 2 times the cutoff in the on-book market.  Similar results are found on other stocks as well.


\subsection{Bootstrapping the largest eigenvalues}


The weaknesses of the parametric eigenvalue test can be remedied by focusing only on the largest eigenvalues and making a bootstrap test.  For each month we construct a realization of the null hypothesis by randomly shuffling buy and sell periods for each institution.  In this way we obtain a bootstrapped strategy matrix $M$ in which the institutions' strategies are uncorrelated, while preserving the number of buying, selling or inactivity periods for each institution.  The shuffling therefore preserves the marginal correlations between strategies, meaning that long term (monthly or longer) correlations between institutions are not altered.  

The shuffling also removes serial correlation in each institution's strategy.  For most institutions this is not a problem because they do not display autocorrelations in the first place.  However, for the group of institutions that do show autocorrelated strategies, this can be an issue.

From the bootstrapped strategy matrix we calculate the correlation matrix and the eigenvalues.   This is repeated for each month separately 1000 times.  

As already noted, instead of looking at the significance of  \emph{all} empirical eigenvalues, we focus only on the largest two eigenvalues for each month.  Correspondingly, we compare them with the null distribution of the two largest eigenvalues for each month: From each of the 1000 simulated correlation matrices, we keep only the largest two.   We are therefore comparing the empirical largest two eigenvalues with an ensemble of largest eigenvalues from the 1000 simulated correlation matrices that correspond to the null appropriate to that month.  In this way the variations of $Q$ and $\sigma$, as well as small sample properties, are taken into account in the test.

\begin{figure*}[hptb]
\begin{center}
	\includegraphics[scale=0.53]{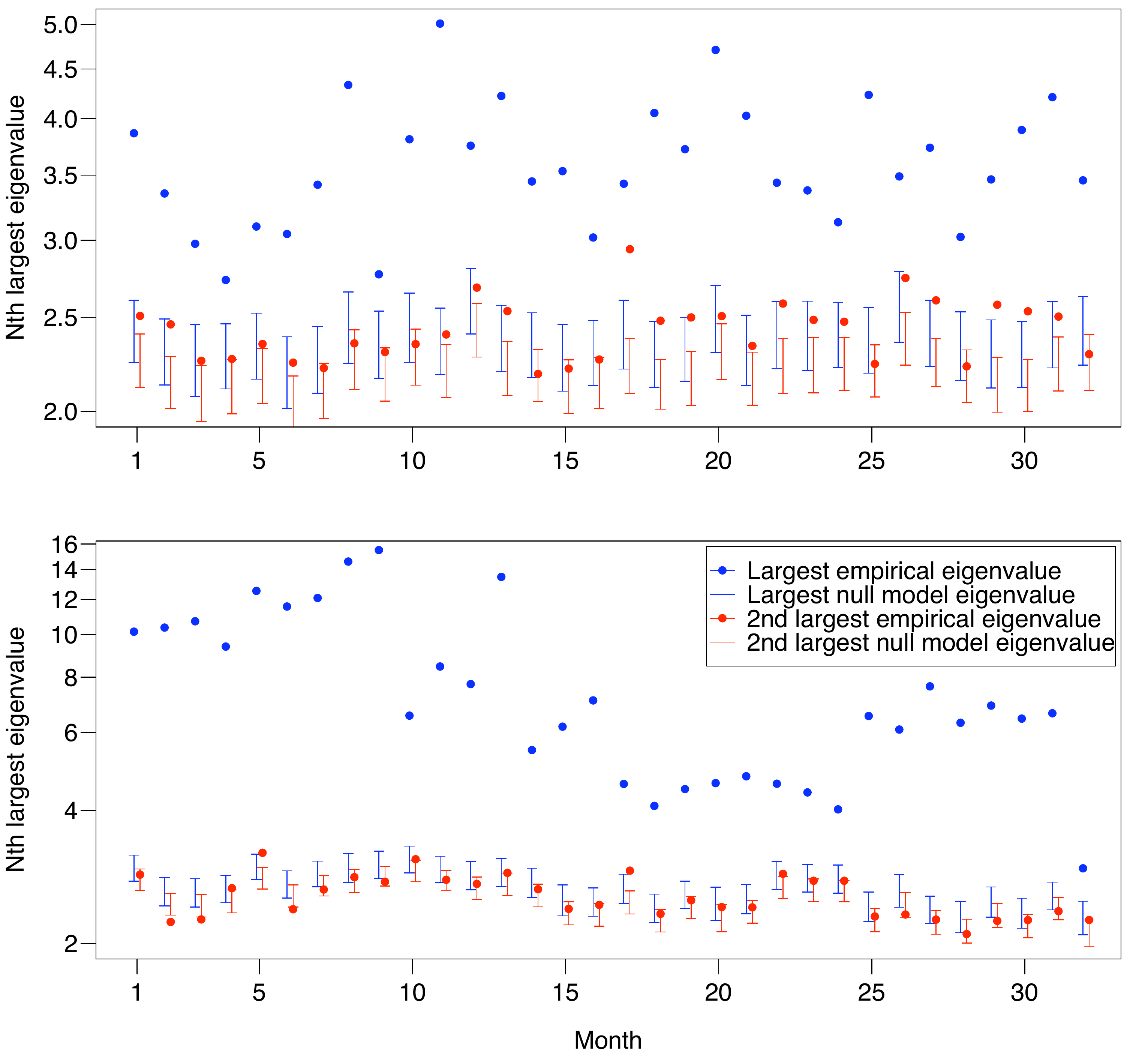}
\caption{Largest eigenvalues of the correlation matrix over the 32 months for the stock Vodafone.  The top figure is for on-book trading, the bottom for off-book trading.  Blue points represent the largest empirical eigenvalues and are to be compared with the blue error bars which denote the null hypothesis of no correlation.  Red points are the second largest eigenvalues and are to be compared with the red error bars.  The error bars are centered at the median and and correspond to two standard deviations of the distribution of largest monthly eigenvalues under the null}
\label{VOD_eigenTimeSeries}
\end{center}
\end{figure*}

Figure ~\ref{VOD_eigenTimeSeries} shows the results for all 32 months of trading in Vodafone.  Again, the top figure shows the monthly eigenvalues for the on-book market while the bottom figure for the off-book market.  The largest empirical monthly eigenvalues are shown as blue points.  They are to be compared with the blue vertical error bars which represent the width of the distribution of the maximum eigenvalues under the null.  The error bars are centered at the median and represent \emph{two} standard deviations of the underlying distribution.  Since the distribution is relatively close to a normal, the width represents about 96\% of the density mass.  The analogous red symbols show the second largest eigenvalue for each month.  We first note that the median of the distribution of the maximum eigenvalue under the random null fluctuates roughly between  2.4 and 2.5.  These values are not so different from $\lambda^{max} = 2.5$ which we used in the parametric test.  It even seems that in small samples and with ternary data, the tendency of $\lambda^{max}$ is to decline as the number of points used decreases, further strengthening the parametric test. The same conclusion can be drawn by looking at the off-book market.  

The largest eigenvalue is significant in all months on both markets.  Corroborating the parametric test,  the largest eigenvalues on the off-book market are relatively further away from the corresponding null than for the on-book market, confirming the observation that the correlations are stronger for off-book trading.  However, being stronger, they are perhaps of a more simple nature:  The second largest eigenvalue is almost never significant for off-book trading, while on the on-book market it is quite often significant.

\subsection{Clustering of trading behaviour}
\label{clustSect}
The existence of significant eigenvalues allows us to use the correlation matrix as a distance measure in the attempt to classify institutions into groups of similar or dissimilar trading patterns.  We apply clustering techniques using a metric chosen so that
two strongly correlated institutions are 'close' and anti-correlated institutions are 'far away'.  A functional form fulfilling this requirement and satisfying the properties of being a metric is ~\cite{Bonanno00}
\begin{equation}
d_{i,j} = \sqrt{2 \cdot (1-\rho_{i,j})}, 
\end{equation}
where $\rho_{i,j}$ is the correlation coefficient between strategies $i$ and $j$.  
We have tried several reasonable modifications to this form but without obvious differences in the results.  Ultimately the choice of this metric is influenced by the fact that it has been successfully used in other studies~\cite{Bonanno00}.  We use complete linkage clustering, in which the distance between two clusters is calculated as the maximum distance between its members.  We also tried using minimum distance (called ``single linkage clustering"), which produced clusters similar to minimal spanning trees but without obvious benefits\footnote{We have also constructed minimal spanning trees from the data but without an obvious interpretation.}.

\begin{figure*}[hptb]
\begin{center}
	\includegraphics[scale=0.8]{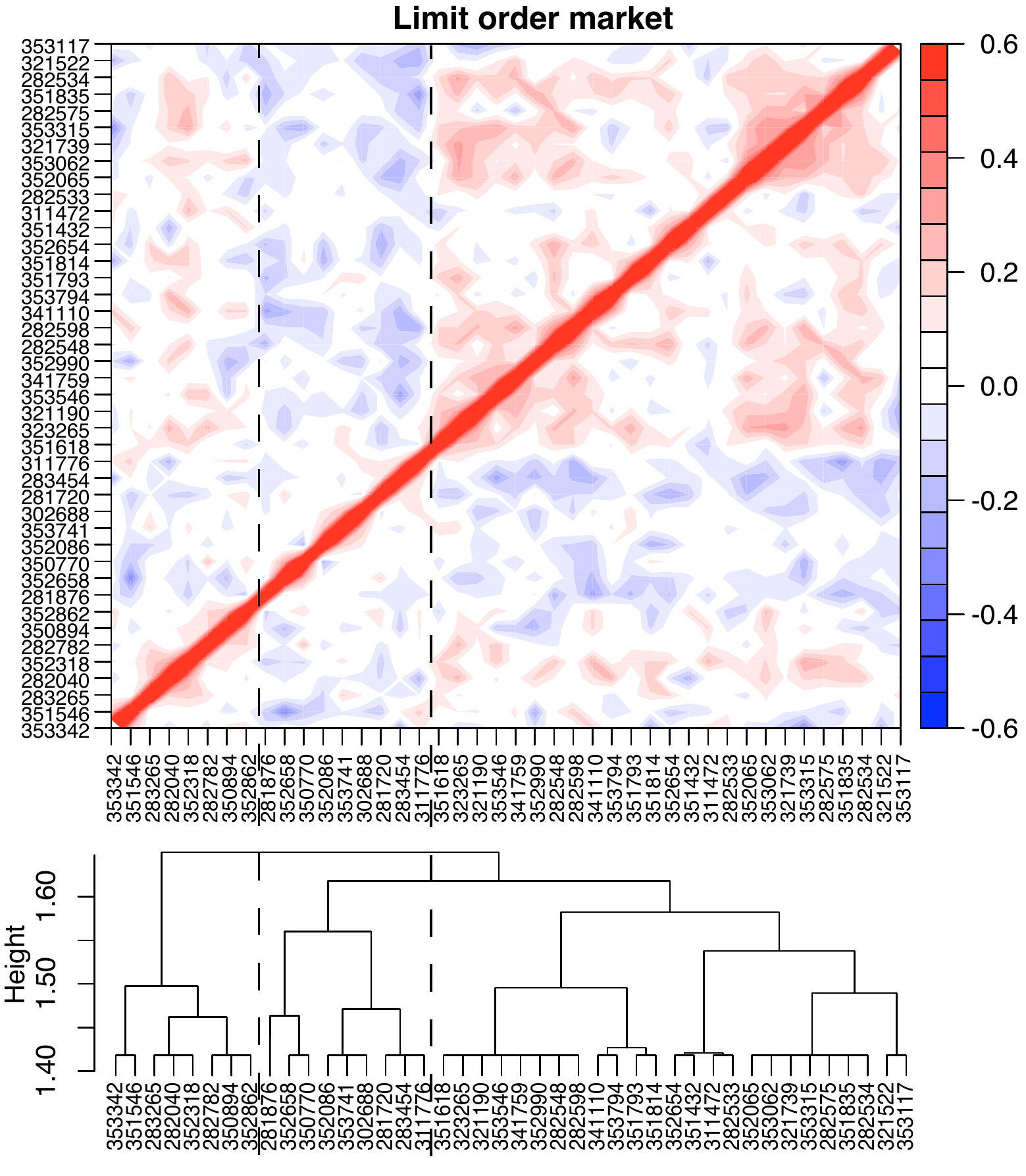}
	\caption{Correlation matrix and the clustering dendrograms for on-book trading in VOD in November 2000.  The correlated and anti-correlated groups of institutions are easily identifiable, however, for this month, the clustering algorithm does not properly classify the institutions at the top clustering level.  We have added lines to help guide the eye to perhaps a better clustering than the algorithm came up with.  It seems that the leftmost group of correlated institutions should have been clustered together with the rest of the correlated institutions.}
\label{bookClust}
\end{center}
\end{figure*}

The first benefit of creating a clustering is to rearrange the columns of the correlation matrix according to cluster membership.  In the top part of figure~\ref{dendrograms} we already showed the rearranged correlation matrix for off-book trading in Vodafone for May 2000.  In the bottom part is the corresponding dendrogram.  In the correlation matrix one notices a highly correlated large group of institutions as the red block of the matrix.  One also notices a smaller number of institutions with strategies that are anti-correlated with the large group.  These institutions in turn are correlated among themselves.  Finally, to the right part of the matrix there is a group of institutions that is weakly correlated with both of the previous two.  These basic observations are also confirmed in the clustering dendrogram - the dendrogram is plotted so that the institutions in the correlation matrix correspond to the institutions in the dendrogram.  Cutting the dendrogram at height 1.7 for example, we recover the two main clusters consisting of the correlated red and the anti-correlated blue institutions.  Cutting the dendrogram at a finer level, say just below 1.6, we also recover the weakly correlated cluster of institutions.  The structure of the dendrogram below 1.4 is suppressed for clarity of the figure, as those levels of detail are noisy.  For other months and other stocks we observe very similar patterns.  The top clustering level typically will classify institutions as a larger correlated group and a smaller anti-correlated group.  

The clustering for the on-book market is similar, though weaker.  Figure~\ref{bookClust}  shows the correlation matrix and the clustering dendrogram for the on-book trading for the same month and stock as the example we showed earlier in figure~\ref{dendrograms}.  We again see correlated and anti-correlated groups of institutions, as well as the weakly correlated group.  The clustering algorithm in this case, however, does not select the correlated and anti-correlated groups at the top level of the clustering, selecting rather the weakly correlated group in one cluster and the other two in the other.  At a finer level of clustering (lower height in the dendrogram) the three groups are clustered separately.  
The clustering algorithm and the distance metric we currently use may not be optimal in selecting the institutions into clusters, but there is indication that the clustering makes sense.   In any case, the existence of clusters of institutions based on the correlation in their strategies suggests that it may be possible to develop a taxonomy of trading strategies.  


\subsection{Time persistence of correlations}
Time persistence, when it is possible to investigate it, offers a fairly robust and strong test for spuriousness.  If a correlation is spurious it is not likely to persist  in time.  In contrast, if the correlations are persistent than the clusters of institutions also persist in time.  As noted before, the LSE rescrambles the codes assigned to the institutions at the turn of each month.  It is therefore not possible to simply track the correlations between institutions in time.    Fortunately, there is a partial solution to this problem.  By exploiting other information in the dataset we are able to unscramble the codes over a few months in a row for some institutions.  Unfortunately, the method works only for trading on the on-book market and typically does not work for institutions that do not trade frequently\footnote{In the LSE data we use each order submitted to the  limit order book is assigned a unique identifier.  This identifier allows us to track an order in the book and all that happens to it during its history.  If at the turn of the month (the scrambling period) an institution has an order sitting in the book, we can connect the institution codes associated with the order before and after the scrambling.  For example, if an order coded AT82F31E13 was submitted to the book on the 31st by institution 2331, and that same order was then canceled on the 1st by institution 4142, we know that the institution that was 2331 was recoded as 4142.  This typically allows us to link the codes for most active institutions for many months in a row, and in several cases even for the entire 32 month period.  The LSE has indicated that they do not mind us doing this, and has since provided us with the information we need to unscramble all the codes.}.
Therefore, the results reported in this section concern only the on-book market and are based mostly on more active institutions.  Since the correlations are typically stronger in the off-book market, we believe the results shown here would hold also for the off-book market, and perhaps be even stronger.
\begin{table*}
\caption{Regression results of equation~\ref{eg:reg} for correlations between institutions for two consecutive months.  Significant slope coefficients show that if two institutions' strategies were correlated in one month, they are likely to be correlated in the next one as well.  The table does not contain the off-book market because we cannot reconstruct institution codes for the off-book market in the same way as we can for the on-book market.   The $\pm$ values are the standard error of the coefficient estimate and the values in the parenthesis are the standard p-values.}
\begin{tabular}{|l ||l |l| l|} 
\hline
\multicolumn{4}{|c|}{On-book market} \\
\hline \hline
{\bf Stock} & {\bf Intercept} & {\bf Slope} & $R^2$\\
\hline 
AAL &  -0.010 $\pm$ 0.004 (0.02) & {\bf 0.25 $\pm$ 0.04 (0.00)} & 0.061 \\
AZN &  {\bf -0.01 $\pm$ 0.003 (0.00)} & {\bf 0.14 $\pm$ 0.03 (0.00)} & 0.019   \\
LLOY &  0.003 $\pm$ 0.003 (0.28) & {\bf 0.23 $\pm$ 0.02 (0.00)} & 0.053 \\
VOD  & {\bf 0.008 $\pm$ 0.001 (0.00)} & {\bf 0.17 $\pm$ 0.01 (0.00)} & 0.029\\
\hline
\end{tabular}
\label{table}
\end{table*}

Given the problems with tracking institutions in time we focus only on persistence up to two months.  To form a dataset we seek all pairs of institution codes that are present at the market for two months in a row.  For all such pairs we compare the correlation in the first of the two months $c_1$ to the correlation in the second of the two months $c_2$.   If the correlation between two institutions was high in the first month, we estimate how likely is it that it will be high in the second month as well by calibrating a simple linear regression
\begin{equation}
\label{eg:reg}
c_2 = \alpha + \beta \cdot c_1 + \epsilon, 
\end{equation}
assuming $\epsilon$ to be i.i.d. Gaussian.  For the stock VOD we identify 7246 linkable consecutive pairs, for AZN 1623, for LLOY 1930 and for AAL 640.  All the regressions are well specified - the residuals are roughly normal and i.i.d.  The regression results for the on-book market are summarized in table~\ref{table}.  All stocks show significant and positive slope coefficients with $R^2$ around 5\%.  Correlated institutions tend to stay correlated, though the relationship is not strong.


Another sign of persistence is 
if an institution gets consistently clustered in a given cluster.  If two institutions tend to be clustered in a given cluster more often than random then we can infer that the cluster is meaningful.  For this purpose we must have a way to distinguish the clusters by some property.  A visual examination of many correlation matrices and dendrograms makes it clear that it is often the case that the two top level clusters are typically of quite different sizes.  It seems natural to call them the \emph{majority} and the \emph{minority} cluster.  Even though it was not always the case, the number of members in the two top clusters differed by a large number more often than not.  Acknowledging that this may not be a very robust distinguishing feature, we choose it as a simple means to distinguish the main clusters.

The probability that an institution would randomly be clustered in the minority a given number of times is analogous to throwing a biased coin the same number of times, with the bias being proportional to the ratio of the sizes of the two clusters.  If the probability for being in the minority was a constant $p$ throughout the $K$ months, the expected number of times $x$ an institution would randomly end up in the minority would be described by a binomial distribution 
\begin{equation}
B(x, p, K) = {x \choose K} p^x (1-p)^{k-x}.  
\end{equation}
In our case, however, the probability of being in the minority is not a constant, but varies monthly with the number of active institutions and the size of the minority.  If the size of the minority is half the total number of institutions, the probability of ending in the minority by chance is 1/2.  If the size of the minority is very small compared to the number of total institutions, the probability of ending in that cluster by chance is consequently very small.  Denoting by $\nu_k$ the number of active institutions in month $k$ and by $\mu_k$ the number of institutions in the minority cluster, then the probability for an institution to be in minority for month  $k$ by chance is $p_k = \mu_k/\nu_k$.  The expected number of times for an institution to be in the minority by chance is then
\begin{equation}
P(x, p_k, K) = \prod_{k \in \mbox{min}} p_k \cdot \prod_{k \in \mbox{maj}} (1-p_k),  
\end{equation}
where $k$ indexes the months in which the institution was in the majority or minority.  
A further complication is that not all institutions are active on the same months, so that the probability density differs from institution to institution: Depending on which months the institution was active, the above product picks out the corresponding probabilities $p_k$.  Because of this complication we calculate the probability density for each institution through a simulation.  We simply pick out the months the institution was active, for each month draw a trial randomly according to $p_k$, and  calculate the number of times the trial was successful, i.e., that the institution ended up in the minority.  Repeating this many times we get the full distribution function for the number of times the institution can end up in the minority at random for each institution.
\begin{table}[htbp]
\begin{center}
\caption{Result of the test on minority members for on-book trading in Vodafone.  In bold are institutions whose behavior is not consistent with the hypothesis of random behavior.}
\begin{tabular}{|c|c c c |} 
\hline
Inst. & Times in & Out of & Prob. of non-\vspace{-0.2cm}  \\
code & minority & possible &-random behavior\\
\hline
{\bf 3265} &  16 & 32 & 0.99 \\
2548 & 7 & 32 & 0.14 \\
2575 & 6 & 32 & 0.07 \\
2533 & 3 & 19 & 0.11 \\
\bf{2040} & 14 & 31 & 0.97 \\
1720 & 9 & 20 & 0.93\\ 
1876 & 5 & 14 &  0.73 \\
2688 & 8 &  30 & 0.34 \\
{\bf 1776} & 11 & 22 &  0.99 \\
2086 & 9 & 23 &  0.86 \\
{\bf 0867} & 10 & 22 &  0.95 \\
{\bf 2995} & 12 & 20 &  1.00 \\
2569 & 7 & 21 &  0.64 \\
\hline
\end{tabular}
\label{minority}
\end{center}
\end{table}

Similarly, because we are using the institution codes over intervals of more than one month, we can perform this test only for institutions on the on-book market.  We limit the test to the stock Vodafone and apply it only on institutions that we can track for more than 12 out of the 32 months.  This results in 13 institutions on which we base the test.  For other stocks we are not able to track institutions for long periods and the power of the test would be weak.

The results for the 13 institutions are given in table~\ref{minority}.   The leftmost column is the institution code, followed by the number of times that institution has been in the minority.  The column named 'Out of possible' counts the number of months an institution has been present in the market - it is the maximum number of times it could have been in the minority.  Finally, the rightmost column gives one minus the probability that the institution could have randomly been so many times in the minority.  
We choose to display one minus the probability as it represents the probability of accepting the hypothesis that the behavior of that institution is not consistent with the random null hypothesis.  Most institutions have quite high probabilities of non-random behavior and in bold we select the institutions which pass the test at the 5\% level. Out of 13 institutions, 5 of them have been in the minority cluster more often than they would have been just by chance at the 5\% acceptance level.  This is substantially higher than the expected number of 0.65 out of 13 tested at this acceptance level.

\section{Conclusions}
We have shown that even very crude definitions of institutions' strategies defined on intervals of an hour period produce significant  and persistent correlations.  On the off-book market these correlations are organized in a way that there is typically a small group of institutions anti-correlated with a larger second group.  The strategies within the two groups are correlated.  Clustering analysis also clearly reveals this structure.  The volume transacted by the smaller group, typically containing no more than 15 institutions on Vodafone, accounts for about half of the total trading volume.  The larger group, typically of around 80 institutions on Vodafone, transacts the remaining half of the total trade volume.  This is an indication that the smaller group can be identified as the group of dealers on the off-book market.  They provide liquidity for the larger group of institutions and their strategies are anti-correlated: the dealers buy when the other institutions are selling and vice versa.  The single large monthly eigenvalue in the off-book market is related to this basic dynamics.

Contrary to the off-book market, the on-book market does not display only one large eigenvalue.  There are typically one or two significant eigenvalues for each month.  The eigenvalues are relatively smaller and the correlations not as strong.  Still, we are able to identify the basic clustering structure seen on the off-book market,  namely a small and large group of anti-correlated strategies.  However, the volume traded by the small cluster does not seem to equal the volume of the large cluster.  The dynamics seems to be more complicated.  The largest eigenvalue may still be related to transactions between the two clusters of institutions, however the occasional second largest eigenvalue suggests that there is more complicated dynamics taking place. 

These results suggest that trading on the LSE is a relatively structured process in the aspect of trading strategies.  At a given time, there are groups of institutions all trading in the same direction, with other groups trading in the opposite direction, providing liquidity.  

It is important to stress that what we have conveniently labeled a ``strategy" is more typically a collection of strategies all being executed by the same member of the exchange.  From this point of view it is particularly remarkable that we observe heterogeneity, as it depends on the tendency of certain types of strategies to execute through particular members of the exchange (or in some cases that pure strategies take the expense to purchase their own membership).  One expects that if we were able to observe actual account level information we would see much cleaner and stronger similarities and differences between strategies.

\bibliographystyle{apalike}  
\bibliography{iiz}

\end{document}